\documentclass[twocolumn,aps,prl,showpacs,preprintnumbers,groupedaddress]{revtex4-1}
\usepackage{amsmath}
\usepackage{amssymb}
\usepackage{dcolumn}
\usepackage{graphicx}
\usepackage{longtable}
\usepackage{bm}
\usepackage{color}
\usepackage{epsfig}

\newcommand{\mysection}[1]{{\it #1.}}

\begin{document}

\title{Possible Half Metallic Antiferromagnetism in a Double Perovskite Material \\
with Strong Spin-Orbit Couplings}
\author{Madhav P. Ghimire, Long-Hua Wu, and Xiao Hu}

\affiliation{International Center for Materials Nanoarchitectonics (WPI-MANA), National Institute for Materials Science,Tsukuba 305-0044, Japan}

\date{\today}

\begin{abstract}
     Using the first-principles density functional approach, we investigate a material Pr$_2$MgIrO$_6$ (PMIO) of double perovskite structure synthesized recently. According to the calculations, PMIO is a magnetic Mott-Hubbard insulator influenced by the cooperative effect of spin-orbit coupling (SOC) and Coulomb interactions of Ir-5$d$ and Pr-4$f$ electrons, as well as the crystal field. When Pr is replaced with Sr gradually, the system exhibits half metallic (HM) states desirable for spintronics applications.
In [Pr$_{2-x}$Sr$_x$MgIrO$_6$]$_2$, HM antiferromagnetism (HMAFM) with zero spin magnetization in the unit cell is obtained for $x=1$, whereas for $x=0.5$ and 1.5 HM ferrimagnetism (HMFiM) is observed with $\mu_{\rm tot}=3\mu_{\rm B}$ and $\mu_{\rm tot}=-3\mu_{\rm B}$ per unit cell respectively. It is emphasized that the large exchange splitting between spin-up and spin-down bands at the Fermi level makes the half metallicity possible even with strong SOC.

\end{abstract}

\maketitle

\mysection{Introduction}--
Half metals (HMs) are a class of materials which are metallic in one spin channel, while insulating in the opposite spin channel due to the asymmetric band structure \cite{groot1,leuken,pickett1,wolf,felser1,katsnelson,xiao1}. HMs can generate spin-polarized currents without any external operation, and thus are very useful for spintronics applications. They have been identified in several groups of materials \cite{pickett2,sarma,min1,xiao2,akai,xiao3,muller,xiao4,coey} 
and huge magnitudes of magneto-resistance have been reported \cite{tokura}.
It was noted, however, that spin-polarized current might be hampered by stray fields which stabilize magnetic domains. This drawback can be overcome in HM antiferromagnets (HMAFM), a subclass of HMs characterized further by zero spin magnetization per unit cell \cite{leuken,xiao1}.
A brand new field coined antiferromagnetic spintronics is emerging towards exploration of novel functionalities of antiferromagnets \cite{bgpark,wang,bibes}.

Ideally HM in a stoichiometric material is a quantum state specified by integer spin magnetization in units of Bohr magneton, where the total number of electrons per unit cell is an integer and all valence bands are fully filled in the insulating spin channel. In reality, however, accurate integer spin magnetization has hardly been achieved. An apparent reason is the quality of a crystal. Another, and more intrinsic, reason may be the existence of spin-orbit coupling (SOC), which is especially important for heavy elements.

Generally speaking, SOC is taken as an unfavorable effect for spintronic applications, in which one wishes to keep the spin moment for information processing and encoding. A naive question then arises whether HM can survive in presence of sizable SOC. In presence of other fields, the answer to this question can be positive. Actually, it is revealed that a topological HMAFM state can be generated by simultaneous application of antiferromagnetic exchange field and alternating electric potential in addition to SOC in a double perovskite structure \cite{liang}.
Generalizing the idea into broader classes of materials is expected to provide a new facet for developing functional materials. A newly synthesized double perovskite material Pr$_2$MgIrO$_6$ (PMIO) \cite{mugavero} then comes into our attention with the unique properties: Pr atoms and Ir atoms carry on opposite spin magnetizations, and oxygen octahedra exhibit large crystal distortion which may induce strong crystal field and help in splitting the spin-up and spin-down bands which are expected to compete with the strong SOC in Ir and Pr atoms.

It is also worth noticing that iridates themselves have been attracting significant interests recently which yield various unconventional phases in cooperation with the strongly-correlated effects due to Coulomb interaction among electrons. For instance, Sr$_2$IrO$_4$ has been evidenced by an experiment as a $J_{\rm {eff}}=1/2$ 
Mott-insulator \cite{kim1,kim2,okada}. 
Ln$_2$Ir$_2$O$_7$ is predicted to transform from a topological band insulator to a topological Mott insulator \cite{pesin}. The topological semimetal state has been predicted in Y$_2$Ir$_2$O$_7$ characterized by Fermi arcs on surface \cite{xiangang}.

We have performed first-principles density-functional-theory (DFT) calculations on PMIO. It is found that PMIO is a Mott-Hubbard insulator, where a Pr atom carries 2$\mu_{\rm B}$ and an Ir atom carries -1$\mu_{\rm B}$ moment, resulting in
$\mu_{\rm tot}=6\mu_{\rm B}$ per unit cell of [Pr$_2$MgIrO$_6$]$_2$.
This material is interesting in the sense that  (i) the topmost valence states close to Fermi level ($E_{\rm F}$) are exclusively spin-up bands contributed from $d$ electrons of Ir, and (ii) the A-site element Pr provides both charge and spin magnetization as opposing to most perovskite materials. Therefore, holes doped into the system tend to exhibit spin-up polarization, which makes this new material a unique platform for material tailor with simultaneous control on charge and spin. Specifically, we consider the replacement of Pr by Sr, with Sr being non-magnetic and donating one electron less than Pr.
We find that the material [Pr$_{2-x}$Sr$_x$MgIrO$_6$]$_2$ is HMFiM with $\mu_{\rm tot}=3\mu_{\rm B}$ and $-3\mu_{\rm B}$ at $x=0.5$ and 1.5 respectively, and HMAFM with $\mu_{\rm tot}=0$ at $x=1$. 
The interplay among Coulomb repulsion, SOC and the crystal field plays an important role in this material. To the best of our knowledge, this is the first prediction of HMAFM with large SOC.

\mysection{Crystal structures and methods}--
 The crystal structure of PMIO is shown in Fig.~1(a) which falls in the space group \emph{P}2$_1/n$ with monoclinic-distortion derived from the double perovskite structure. It has structural distortions due to the tilting and rotation of IrO$_6$ octahedron in addition to the different bond-lengths between Ir and oxygen atoms.

\begin{figure}[ht]
\centering
\psfig{figure=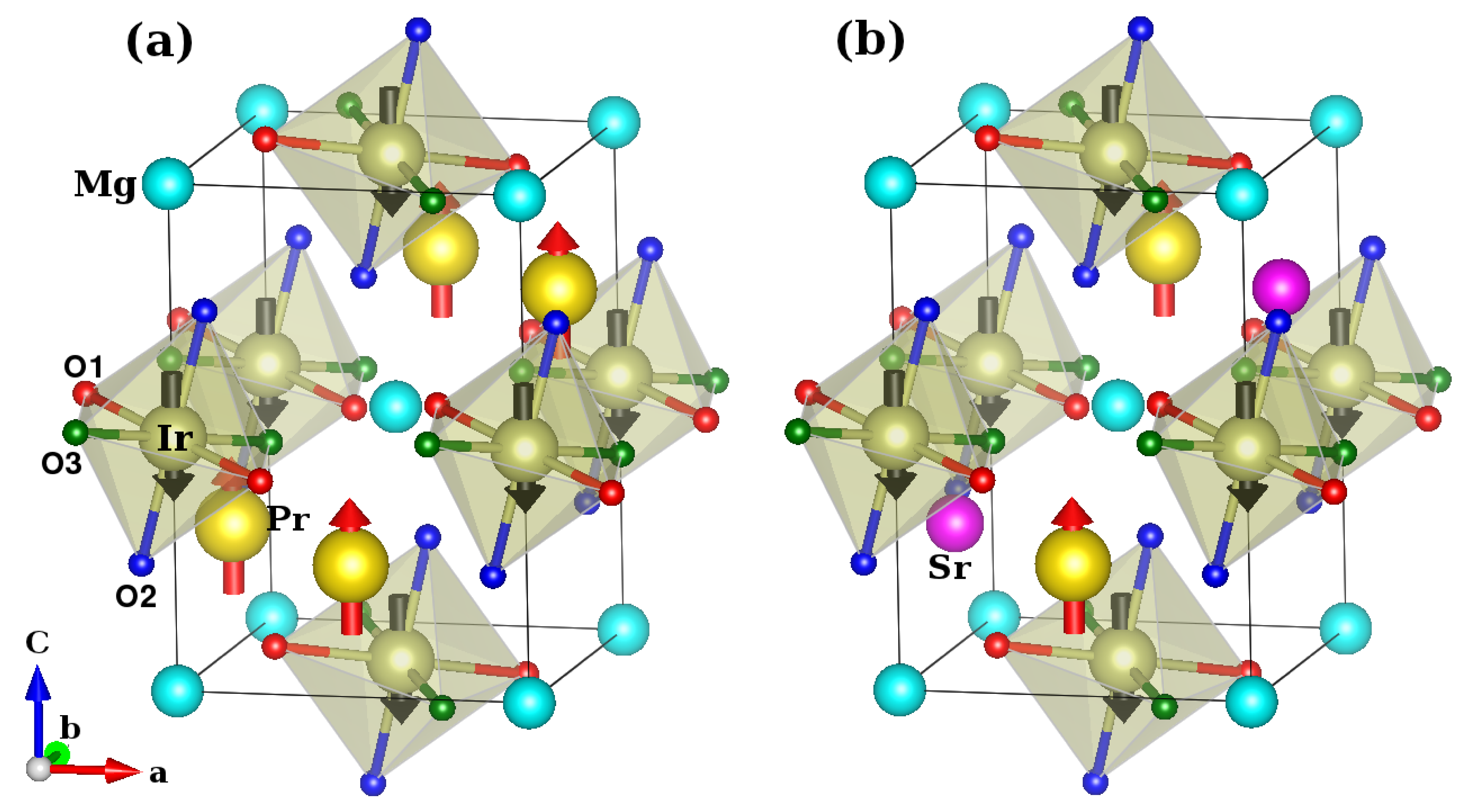,width=8.5cm}
\caption{Crystals of double perovskite structure: (a) parent material [Pr$_2$MgIrO$_6$]$_2$; (b) material with half Pr replacement [PrSrMgIrO$_6$]$_2$.
    The red (black) arrows indicate the direction of Pr (Ir) spins along the $c$ direction which is the easy axis.}
   \label{fig:1}
\end{figure}

In order to provide a realistic description of the electronic and magnetic structures, a set of first-principles DFT calculations were performed using the full-potential linearized augmented plane wave plus local orbital method implemented in the WIEN2k code \cite{blaha}.
The atomic sphere radii $R_{\rm MT}$ were 2.44, 2.2, 1.91, 2.07 and 1.7 Bohr for Pr, Sr, Mg, Ir and O respectively. A set of 2000 $k$-points were used in the full Brillouin zone. The standard generalized-gradient approximation (GGA) exchange-correlation potential within the PBE-scheme \cite{perdew} were used with Coulomb interaction $U$ of 5eV for Pr and 1.25eV for Ir, respectively \cite{anisimov,note}.
Spin-orbit coupling is considered via a second variational step using the scalar-relativistic eigenfunctions as basis \cite{kunes}.
Starting from the lattice parameters given by experiments \cite{mugavero}, we relax the lattice and reach the stable structure using the VASP package \cite{kresse} with the force convergence set at 0.01 eV/$\rm \AA$.  
 Details of the calculation methods and other results are summarized in supplemental material \cite{supp}.

\begin{figure}[ht]
\centering
\psfig{figure=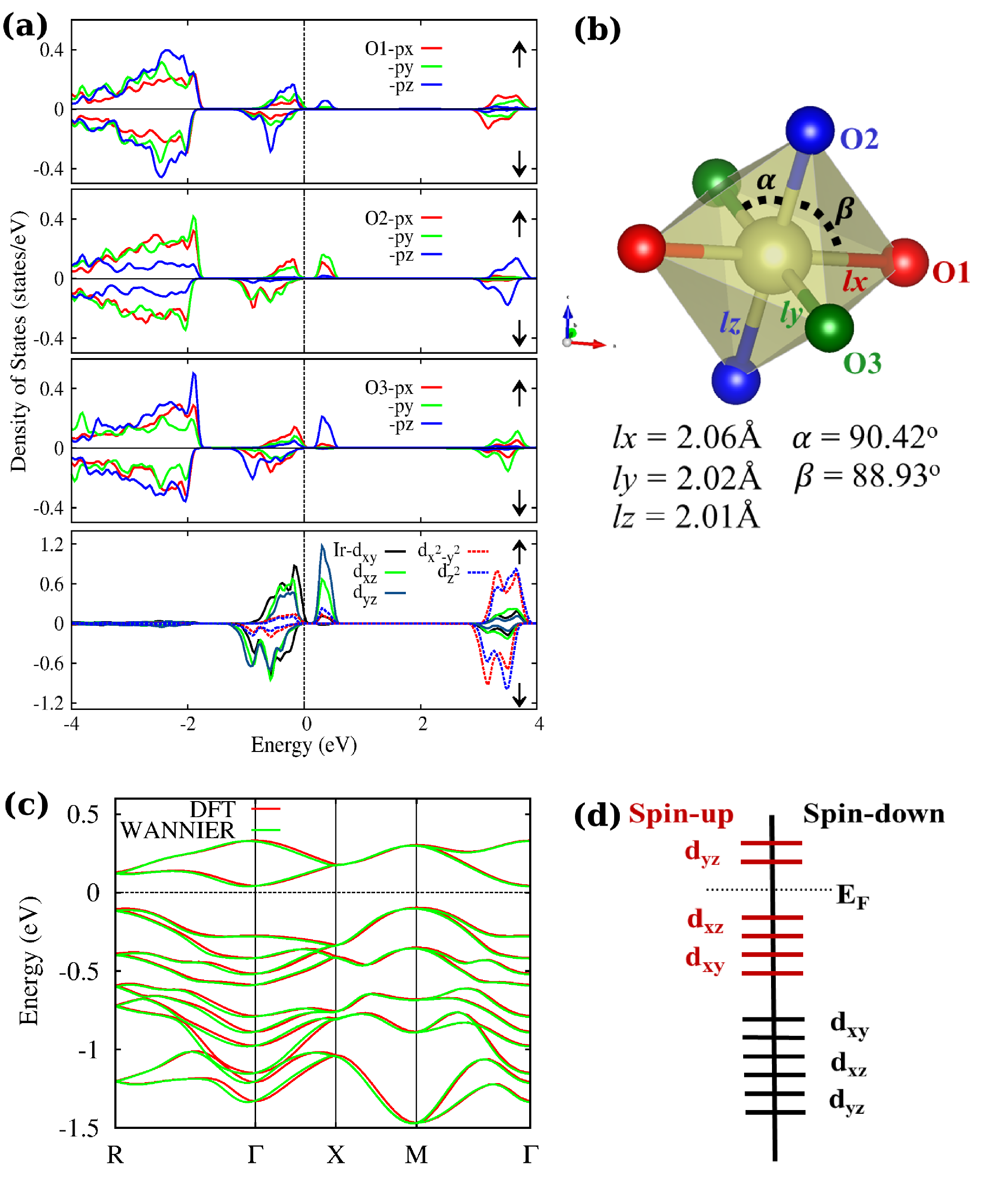,width=9cm}
\caption{(a) Partial density of states for three in-equivalent oxygen-2$p$ states and Ir-5$d$ states in spin-up ($\uparrow$) and spin-down ($\downarrow$) channels, (b) a distorted IrO$_6$ octahedron, (c) band structure  with red (green) curves for results based on DFT (Wannier) method, and (d) schematic band diagram for the ordering of Ir-$t_{2g}$ electrons for the parent material [Pr$_2$MgIrO$_6$]$_2$.}
   \label{fig:2}
\end{figure}

\mysection{Parent material {\rm [Pr$_2$MgIrO$_6$]$_2$}}-- 
  Same as most A-site elements in perovskite materials, in PMIO Pr provides charge to the system and nominally takes the charge state $+3$ with $4f^2$ configuration lying deeply in the valence band. Unlike other cases, however, Pr is in a high-spin state due to strong Hund's coupling. The transition element Ir nominally takes the charge state of $+4$ with $5d^5$ configuration, where five of the totally six $t_{2g}$ orbits are occupied and lie at the top of the valence band, forming a low-spin state due to large crystal field from oxygen octahedron. According to first-principles calculations, there is an energy gap of $\sim0.2$eV at $E_{\rm F}$ (see Fig.~2),
indicating clearly that PMIO is a Mott-Hubbard insulator.

As revealed by the partial density of states (PDOS) in spin-up and spin-down channels \cite{lee} and the band structures obtained from first-principles calculations as well as Wannier downfolding analyses (Fig.~2),
the $t_{2g}$ orbits splits into $d_{yz}>d_{xz}\ge d_{xy}$
in the order of energy, and a large exchange energy $\sim 1.2$eV pushes those in the spin-down channel down away from $E_{\rm F}$. This is caused by distortions of octahedron in the present material, where there are three sorts of oxygen positions with different Ir-O bond lengths $lz\le ly<lx$.

The magnetic property of PMIO is of particular interests. At the ground state obtained from first-principles calculations, Pr couples antiferromagnetically with Ir. The calculated total angular momentum ($\mu_{\rm tot}$) is 6.02$\mu_{\rm B}$ per unit cell (see Table~I).  In an ionic picture, each Pr ion carries moment $+2\mu_{\rm B}$ while Ir ion carries $-1\mu_{\rm B}$, giving rise to $\mu_{\rm tot}$ $=4\times(+2\mu_{\rm B})+2\times(-1\mu_{\rm B})=6\mu_{\rm B}$ in a unit cell, consistent with the first-principles calculations.

\begin{figure}[t]
     \centering
     \psfig{figure=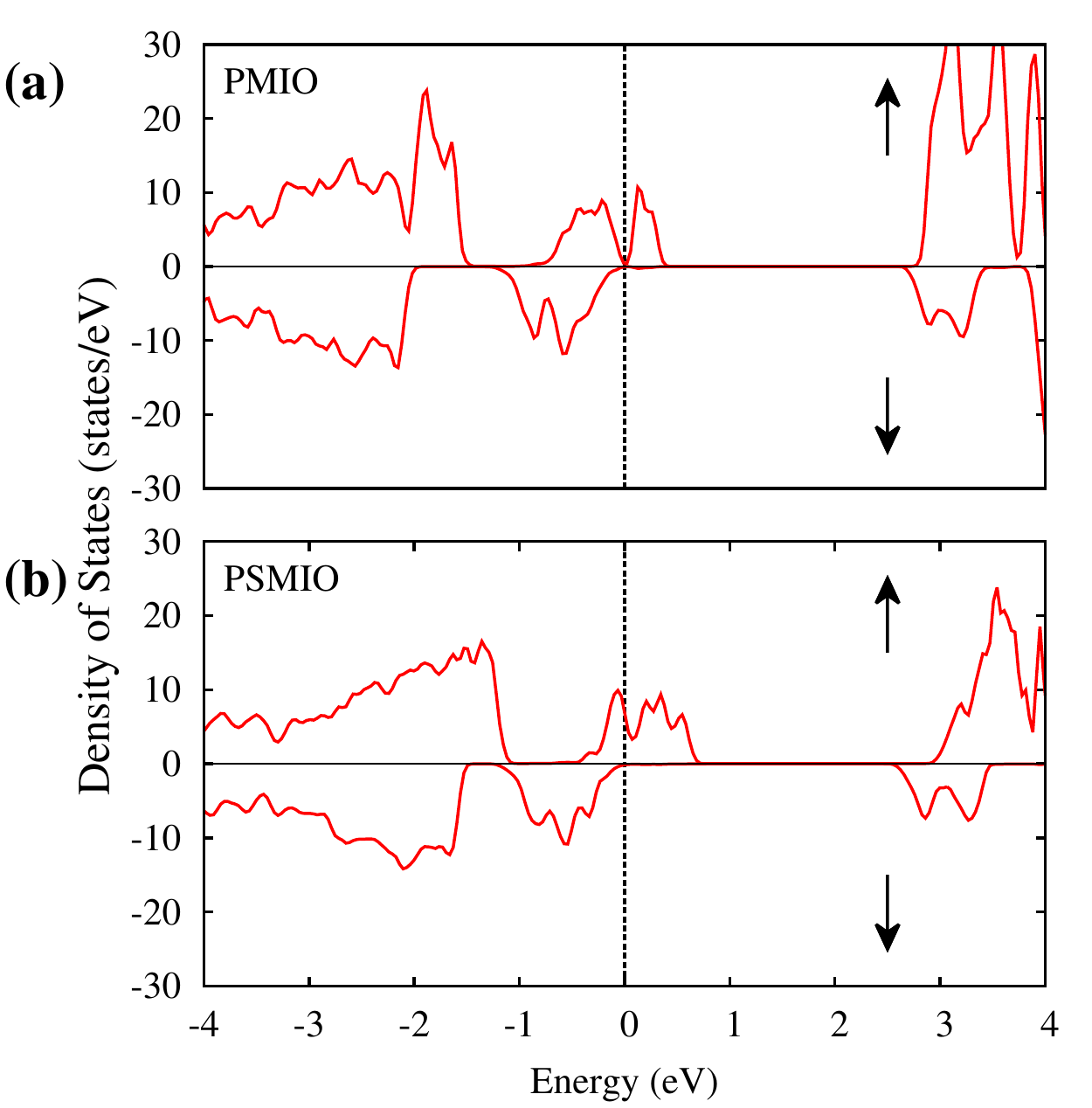,width=\linewidth,height=3.3in}
     \caption{Total density of states obtained by  GGA$+U+$SOC scheme for spin-up ($\uparrow$) and spin-down ($\downarrow$) channels: (a) parent material [Pr$_{2}$MgIrO$_6$]$_2$, (b) material with half Pr replacement [PrSrMgIrO$_6$]$_2$.}
     \label{DOS}
\end{figure}

\mysection{Doped materials {\rm [Pr$_{2-x}$Sr$_x$MgIrO$_6$]$_2$}}--
The above properties of PMIO make it a promising candidate for exploring possible HM states. To be specific, we consider the A-site modification by replacing Pr with Sr, a non-magnetic element usually of $+2$ charge state, which corresponds to hole doping into the parent material. Presuming that the overall magnetic configuration will not be changed upon replacement, a hole will go to the spin-up channel and reduce the total angular moment by $\mu=3\mu_{\rm B}$, with $2\mu_{\rm B}$ taken away by the replaced Pr atom and $1\mu_{\rm B}$ due to $E_{\rm F}$ shift. In this way, one can modify the material with fine control on both spin and charge.

We perform first-principles calculations to check the above idea. Let us focus on the most interesting case of replacement rate $x = 1$, where two Pr atoms are replaced by two Sr atoms, and thus two holes are doped into the system per unit cell. From the view of rigid band model, doping hole shifts $E_{\rm F}$ downward to valence region such that it crosses the topmost occupied states. As revealed by first-principles calculations, the two holes from Sr go to the spin-up Ir-$d_{xz}$ states that lie just below $E_{\rm F}$ in the parent material, and push them up above $E_{\rm F}$. A gap cannot be opened at $E_{\rm F}$ in this case, since the energy difference between $d_{xz}$ and $d_{xy}$ is small due to the almost equal bond lengths $ly$ and $lz$ (see Fig.~2(b)). The spin-down channel remains insulating since no change occurs in the valence states. With spin-up channel metallic and spin-down channel insulating, the system turns to a HM as clearly seen in Fig.~3(b).

As summarized in Table~I, two replaced Pr atoms take away $\mu\simeq 4\mu_{\rm B}$, and the shift of $E_{\rm F}$ associated doped two
holes in the spin-up channel contributes a reduction of $\mu\simeq 2\mu_{\rm B}$ further, which reduces the total angular moment to zero.
These features can also be seen from the spin-density isosurface plot in Fig.~4. 
With the zero total angular moment and HM property, we conclude that the material PrSrMgIrO$_6$
should be a HMAFM. To the best of our knowledge, this is the first proposal for HMAFM with large SOC taken into account.

 \begin{table}[t]
  \centering
  \caption{Moments per atom of Pr and Ir, one set of three in-equivalent oxygen atoms and unit cell ($\mu_{\rm tot}$)
   for replacement rate $x$ in [Pr$_{2-x}$Sr$_x$MgIrO$_6]_2$ from first-principles calculations. The unit of moments is the Bohr magneton $\mu_{\rm B}$. The contributions from individual atoms are within muffin-tins while the total angular moment includes those from interstitial regime.}
  \begin{ruledtabular}
  \begin{tabular}{r c c c c }
{ $x$} &    Pr    &  Ir    &  O &  {$\mu_{\rm tot}$} \\\cline{1-5}
     0.0  & 1.98   & -0.55  & -0.22   & 6.02 \\
   0.5  & 1.97   & -0.76   & -0.32   & 3.07 \\
   1  & 1.96   & -0.98   & -0.43   & 0.08 \\
   1.5  & 1.96   & -1.17   & -0.58   & -2.97 \\
  \end{tabular}
  \end{ruledtabular}
  \label{tab:1}
 \end{table}

HM states are also obtained for the replacement rates $x=0.5$ and $x=1.5$ in [Pr$_{2-x}$Sr$_x$MgIrO$_6$]$_2$ where one and/or three Pr atoms are replaced by Sr atoms per unit cell \cite{supp}. For $x=0.5$, one hole from Sr goes to spin-up Ir-$d_{xz}$ band that was lying in the topmost valence region below $E_{\rm F}$ in the parent material. As a result, Ir-$d_{xz}$ band shifts to conduction region and forms a continuous band with the Ir-$d$ bands in the valence region. This gives rise to metallic state for spin-up channel, while valence states in spin-down channel remain far from $E_{\rm F}$. Similar results have been achieved for $x=1.5$, except that three holes from Sr are transferred to spin-up Ir-$5d$ states.
The two materials are HMFiM with
$\mu_{\rm tot}=+3\mu_{\rm B}$ and $\mu_{\rm tot}=-3\mu_{\rm B}$ respectively.

\begin{figure}[t]
    \centering
    \psfig{figure=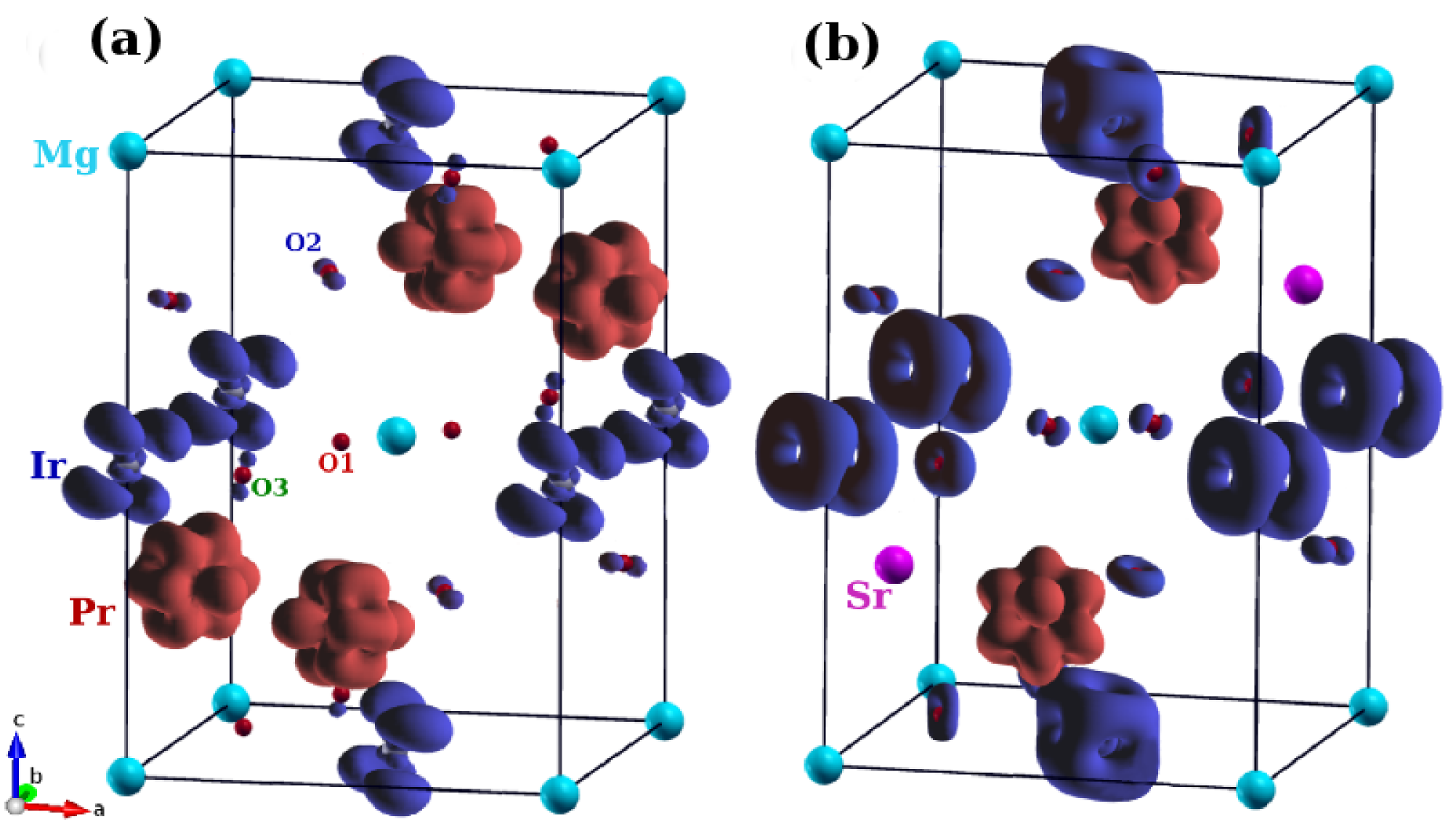,width=\linewidth,height=1.8in}
    \caption{Isosurface of spin magnetization density at $\pm$0.21 $e/$\AA$^3$ with red (blue) for spin up (down): (a) parent material [Pr$_2$MgIrO$_6$]$_2$, (b) material with half Pr replacement [PrSrMgIrO$_6$]$_2$.}
    \label{isosurface}
\end{figure}

\mysection{Discussions}--
In PMIO, SOC is crucially important due to the presence of heavy elements such as Pr and Ir. The orbital moments obtained from first-principles calculations for Pr ($-0.4\mu_{\rm B}$) and Ir ($-0.2\mu_{\rm B}$) are in accordance with the Hund's third rule, where Pr with less than half-filled $f$-shell has its orbital moment anti-parallel to its spin moment, whereas Ir with more than half-filled $t_{2g}$-shell has its orbital moment parallel to its spin moment \cite{kittel}. Hence spin-moment increases for Pr while decreases for Ir to keep the total angular momentum ($\sim6.0\mu_{\rm B}$) unchanged. The angular moments, which are the summation of  spin and orbital moments from individual atom are summarized in Table~I.

Charge-transfer effect \cite{zaanen} is prominent between Ir and oxygen, especially in O2 and O3 due to their shorter bond-lengths with Ir. Charge transfer between O2 and Ir occurs via $p_x$, $p_y$ and $d_{xz}$, $d_{yz}$ states, and between O3 and Ir via $p_z$ and $d_{xz}$, $d_{yz}$ states.
Therefore, O2 and O3 get spin-polarized in parallel with the Ir ions, consistent with the isosurface plot shown in Fig.~4(a).
Doping holes enhances the spin moment further at Ir and oxygen (see Fig.~4(b)). Similar features were reported for Sr-doped LaCoO$_3$ \cite{medling}. 
First-principles calculations on magnetic anisotropy energy indicates that the $c$ axis of the crystal as the easy axis (see Fig.~1) with anisotropy energy of $\sim 27$meV per unit cell
for the parent material.

It is well known that SOC mixes spin-up and spin-down bands, and thus spin-polarization at $E_{\rm F}$ may be affected. It is not the case for the present materials, where the exchange splitting between the spin-up and spin-down DOS at $E_{\rm F}$ is large (see Fig.~3) and the mixing does not happen despite of the strong SOC ($\sim0.35$eV) \cite{matt}. This is the first theoretical prediction of HMAFM based on first-principles calculations involving strong SOC.

In order to check the robustness of half metallicity in the present materials, we consider the disorder effects. There are two main types of disorders, known as antisite disorder \cite{onur} where the positions of Mg and Ir atoms at B and B$'$ sites are interchanged, and cation disorder where Sr atoms replace Pr atoms at different A-site positions. We have confirmed that the HMAFM remains stable in the disordered configurations \cite{supp}.

In the present work, HMAFM and HMFiM have been derived from the same parent material. Thus, using them in an integrated system, one can construct a useful device for spintronics applications without suffering from the problem of lattice mismatching.

\mysection{Conclusions}-- 
Based on the first-principles density functional approach, we propose material tailoring on a Mott-Hubbard insulator Pr$_2$MgIrO$_6$ with double perovskite structure exploiting the cooperative effect from Coulomb interaction, spin-orbit coupling and the crystal field. It is demonstrated that doping holes into the system by replacing Pr with Sr, one can achieve several half metals. Especially, PrSrMgIrO$_6$ is found to be a half metallic antiferromagnet, namely half metal with zero spin magnetization per unit cell, which is ideal for spintronics. It is emphasized that the large exchange splitting between spin-up and spin-down bands at the Fermi level retains the half metallicity even in presence of strong spin-orbit coupling.

    The authors thank R. Yu for valuable discussions. This work was supported by WPI Initiative on Materials Nanoarchitectonics, MEXT, Japan.


\onecolumngrid
\appendix*
\section{Supplementary material: Possible Half Metallic Antiferromagnetism in a Double Perovskite Material \\
with Strong Spin-Orbit Couplings}
\setcounter{figure}{0}
\setcounter{table}{0}
\renewcommand{\thefigure}{S\arabic{figure}}
\renewcommand{\thetable}{S\arabic{table}}

\mysection{VASP calculations}--
We have performed the density functional calculations by using Vienna \textit{Ab initio} Simulation Package (VASP)\cite{vasp} in
addition to WIEN2k \cite{wien}. The generalized gradient approximation (GGA) in the parametrization of Perdew, Burke and Ernzerhof (PBE) \cite{pbe} is used for exchange-correlation potential. The on-site Coulomb interactions are
treated by Dudarev's method \cite{Ueff} with effective $U$ values 5eV for Pr-4$f$ electrons and 1.25eV for the Ir-5$d$ electrons. We use an $8\times8\times8$ $k$-point mesh within the Monkhorst-Pack
scheme \cite{MP} with energy cutoff 400eV. 

\mysection{Downfolding by Wannier functions}--
To get intuitive real-space picture on how Ir-$t_{2g}$ electrons close to Fermi level hop, we
project the bands obtained from density functional theory (DFT) calculations to the localized Wannier functions, i.e. the Ir- $d_{xy}$,
$d_{xz}$ and $d_{yz}$ orbitals. Since there are two Ir atoms in a unit cell, the total number of Wannier functions is
twelve for the two spin channels. 
The hopping integrals of $t_{2g}$ electrons within a same Ir atom
are given in Table \ref{tab:hopping}, while those within the other Ir atom are not listed since they share the
same set of parameters. The hopping integrals between two Ir atoms are almost negligible. As displayed in Table
\ref{tab:hopping}, large energy differences between spin-up and spin-down electrons prevent spin-down electrons from
being pulled up to Fermi level by spin-orbit coupling (SOC). 
Effective on-site energies for all the Wannier functions  obtained are arranged in the order shown in Fig.~2(d) in the
main text. With the dominant nearest-neighbor hoppings among the Wannier orbitals, the DFT result is well reproduced
(see Fig. 2(c) in the main text).

  \mysection{Partial density of states for {\rm [PrSrMgIrO$_6$]$_2$}}--
Fig.~S1 shows the partial DOS contribution from Pr-4$f$, Ir-5$d$ and oxygen-2$p$ states. It is clear that in spin-up channel the Ir-$t_{2g}$ orbits originally in the valence band are now pushed up to cross Fermi level and form a continuous band there. Oxygen $p$ orbits also appear around Fermi level due to the hybridization with Ir $d$ electrons.

\begin{table}[ht]
  \caption{Transfer hopping integrals in units of meV among $t_{2g}$ orbitals along direction ($i$, $j$, $k$) for one of
  the two Ir atoms. For example, the hopping integrals in direction $(1,0,0)$ denote the rates of Ir-$d$ electrons hopping to the Ir atom at the same position but in the nearest unit cell along $x$-axis.
The row order of $6\times 6$ transfer matrix is identical to that of the column. We only show upper
triangular part of self-adjoint transfer matrix.}
  \label{tab:hopping}
    \centering
    \begin{tabular}{c|rrrrrr||c|rrrrrr}
      \hline\hline 
      Direction & $d_{xz},\uparrow$ & $d_{yz},\uparrow$ & $d_{xy},\uparrow$ & $d_{xz},\downarrow$ & $d_{yz},\downarrow$ &
      $d_{xy},\downarrow$ & Direction & 
      $d_{xz},\uparrow$ & $d_{yz},\uparrow$ & $d_{xy},\uparrow$ & $d_{xz},\downarrow$ & $d_{yz},\downarrow$ &
      $d_{xy},\downarrow$  \\\hline
      0,0,0       &  6600     & -209+160$i$ & 73-11$i$    & -5+9$i$ & 31+58$i$& -160+61$i$  &
      0,0,1       &  -10      &    2      &   -2     &   0     &   0     & 0 \\
                &  &  6726       & -83-81$i$   &-27-83$i$&-3+12$i$&  62+127$i$  &
                &          &    6      &    3     &   0     &   0     & 0 \\
                &   &    &  6435       &152-58$i$& -48-130$i$ &  0    &
                &         &          &    4     &   0     &   0     & 0 \\
                &    &  &  & 6085    &141-135$i$ & 54+6$i$     &
                &          &          &         &  -10    &   2     & -2   \\
                &   &    &    &     & 6031   &  43+75$i$    &
                &          &          &         &        &   5     & 2  \\
                &      &     &         &      &       & 6190 &    
                &          &          &         &       &        & 3  \\ \hline
      1,0,0       &  -7       &   -8      &  -19     &   0     &   0     & 0  &
      1,1,0       &  -8       &     0     &    6     &   0     &  0      &  0 \\
                &         &   33      &  -10     &   0     &   0     & 0  &
                &          &     4     &    6     &   0     &  0      &  0 \\
                &        &        &  -64     &   0     &   0     & 0  &
                &          &          &    0     &   0     &  0      &  0 \\
                &          &          &         &  -1     &   0     & -20  & 
                &         &        &         &  -11    &  0      &  7 \\ 
                &          &          &         &        &   36    & -8  & 
                &         &        &         &        &  3      &  5 \\
                &          &          &         &      &       &  -68 &
                &         &        &         &        &        &  1 \\ \hline
      0,1,0       &  -18      &    4      &   37     &   0     &   0     & 0 \\
                &          &   37      &  -28     &   0     &   0     & 0 \\
                &         &        &  -81     &   0     &   0     & 0 \\
                &          &          &         &  -17    &   4     &  39  \\
                &          &          &         &        &   44    & -33 \\
                &          &          &         &       &      & -76  \\ 
      \hline\hline
    \end{tabular}
\end{table}

\begin{figure*}[ht]
  \centering
  \psfig{figure=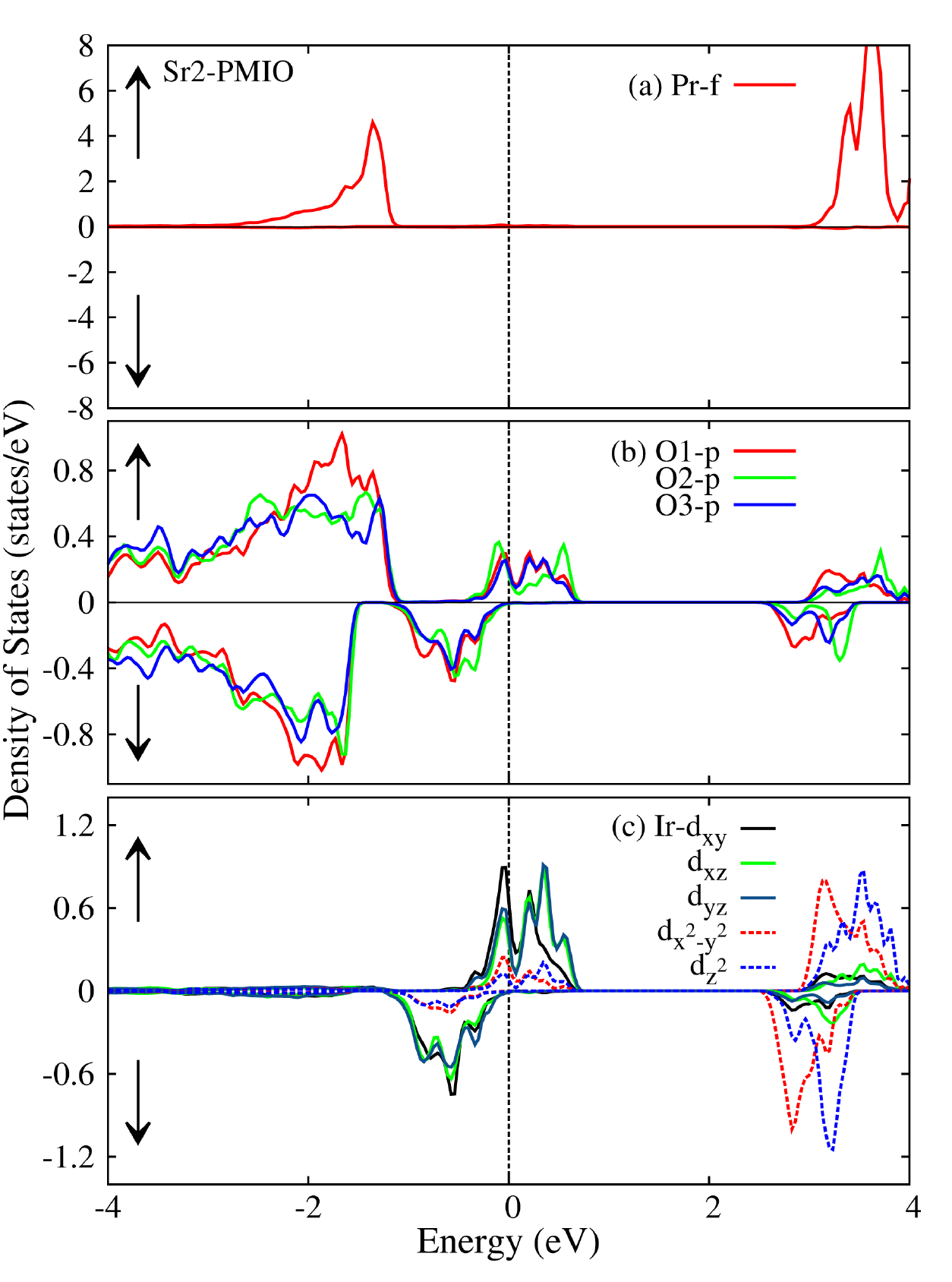,width=3.7in,height=5.in}
  \caption{Partial density of states for Pr-4$f$ states, three in-equivalent oxygen-2$p$ states and Ir-5$d$ states in spin-up ($\uparrow$) and spin-down ($\downarrow$) channels for material with half Pr replacement [PrSrMgIrO$_6$]$_2$ (Sr2-PMIO).}
  \label{Fig. S1}
\end{figure*}

\clearpage
  \mysection{Half metallicity in {\rm [Pr$_{2-x}$Sr$_x$MgIrO$_6$]$_2$}}--
  Half metallic ferrimagnetic states are obtained for the replacement rates $x=0.5$ and $x=1.5$ in
  [Pr$_{2-x}$Sr$_x$MgIrO$_6$]$_2$ where one and/or three Pr atoms are replaced by Sr atoms per unit cell. Shown in
  Fig.~S2 are the total DOS for replacement rates $x=0.5$ and $x=1.5$ respectively. The topmost Ir-$5d$ bands in the valence region of spin-up channel shifts to the conduction region due to doping effects and forms a continuous band with the valence region, resulting in a metallic state for spin-up channel. Spin-down channel remains insulating. The materials are thus a half metals.

\begin{figure*}[ht]
  \centering
  \psfig{figure=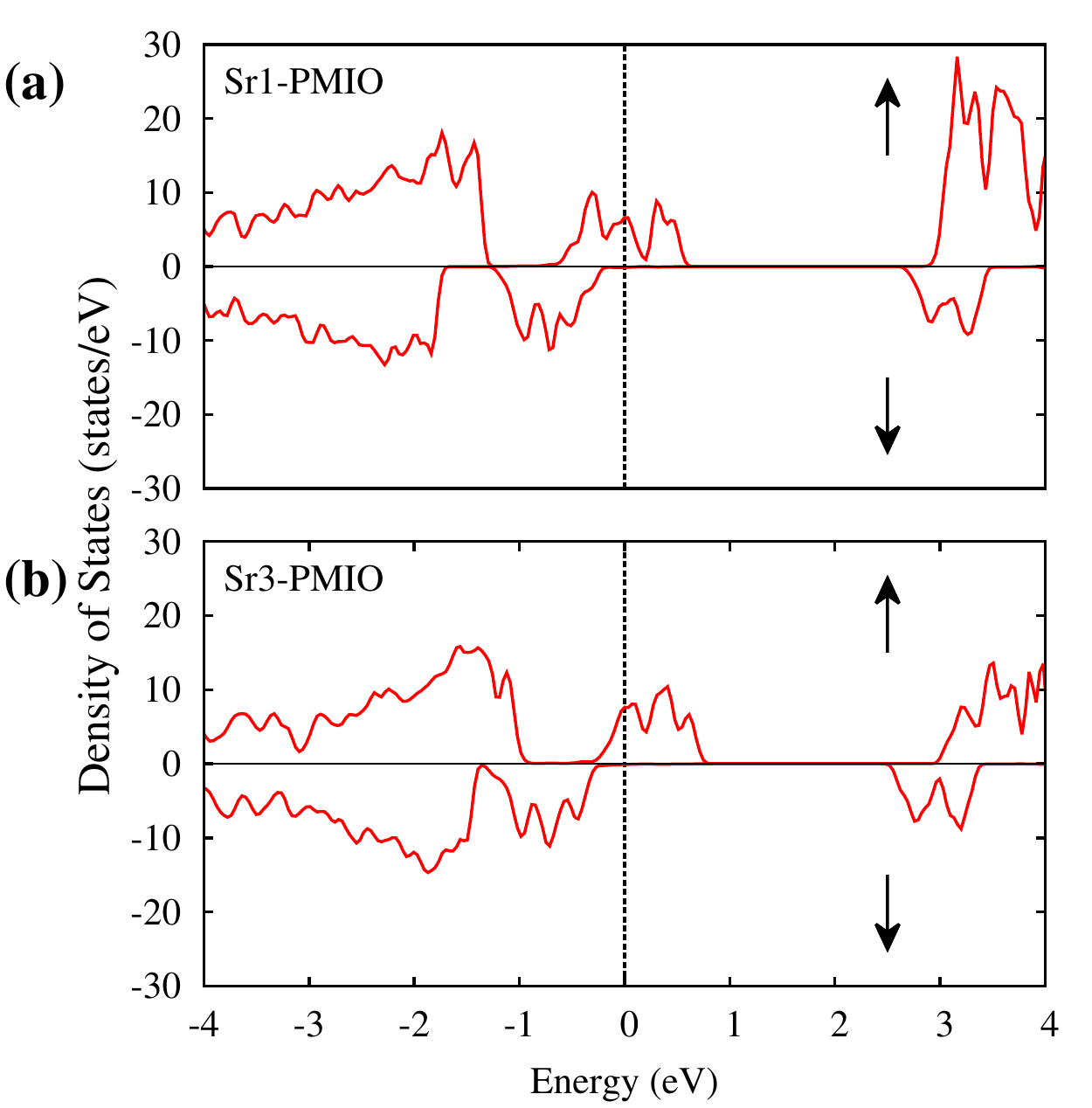,width=4.in,height=5.in}
  \caption{Total density of states obtained for (a) $x=0.5$ (Sr1-PMIO), and (b) $x=1.5$ (Sr3-PMIO) in [Pr$_{2-x}$Sr$_x$MgIrO$_6$]$_2$ in spin-up ($\uparrow$) and spin-down ($\downarrow$) channels.}
  \label{Fig. S2}
\end{figure*}
\clearpage
    \mysection{Robustness of half metallicity in {\rm [PrSrMgIrO$_6$]$_2$}}--
     Shown in Fig.~S3 are two examples of antisite disorder (b), and cation disorder (c) for the replacement rate $x=1$, which are the lowest excited states to the ground state with excitation energy of 400meV and 70meV respectively. The half-metallic states are fully preserved.

\begin{figure*}[ht]
    \centering
    \psfig{figure=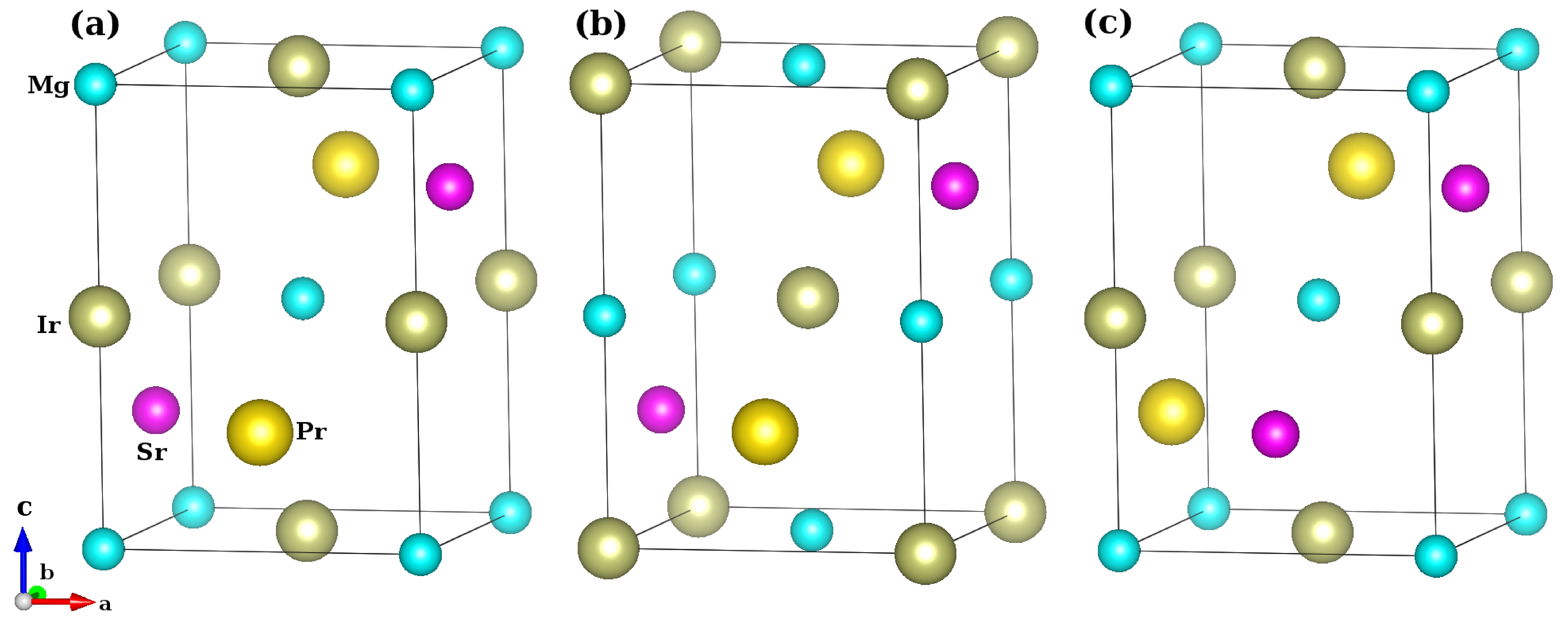,width=\linewidth}
    \caption{Configuration of ground state (a), lowest-excitation state in antisite disorder (b), and cation disorder (c) for the material [PrSrMgIrO$_6$]$_2$. Oxygen atoms are not shown for clear view.}
    \label{Fig.S3}
\end{figure*}


\begin{thebibliography}{99}
  \bibitem{groot1}
     R. A. de Groot, F. M. Mueller, P. G. van Engen, and K. H. J. Buschow, Phys. Rev. Lett. {\bf 50}, 2024 (1983). 
  \bibitem{leuken}
     H. van Leuken, and R. A. de Groot, Phys. Rev. Lett. {\bf 74}, 1171 (1995). 
       \bibitem{pickett1}
     W. E. Pickett, and J. S. Moodera, Physics Today {\bf 54}, 39 (2001). 
  \bibitem{wolf}
     S. A. Wolf, D. D. Awschalom, R. A. Buhrman, J. M. Daughton, S. von Molnar, M. L. Roukes, A. Y. Chtchelkanova, and D. M. Treger, Science {\bf 294}, 1488 (2001). 
  \bibitem{felser1}
     C. Felser, G. H. Fecher, and B. Balke, Angew. Chem. Int. Ed. {\bf 46}, 668 (2007).
  \bibitem{katsnelson}
 	M. I. Katsnelson, Y. Yu. Irkhin, L. Chioncel, A. I. Lichtenstein, and R. A. de Groot, Rev. Mod. Phys. {\bf 80}, 315 (2008).
  \bibitem{xiao1}
     X. Hu, Adv. Mater. {\bf 24}, 294 (2012). 
  \bibitem{pickett2}
     W. E. Pickett, Phys. Rev. B {\bf 57}, 10613 (1998).
  \bibitem{sarma}
     D. D. Sarma, P. Mahadevan, T. Saha-Dasgupta, S. Ray, and A. Kumar, Phys. Rev. Lett. {\bf 85}, 2549 (2000).
  \bibitem{min1}
     J. H. Park, S. K. Kwon, and B. I. Min, Phys. Rev. B {\bf 65}, 174401 (2002).
  \bibitem{xiao2}
     X. Wan, M. Kohno, and X. Hu, Phys. Rev. Lett. {\bf 94}, 087205 (2005); $ibid.$ {\bf 95}, 146602 (2005). 
  \bibitem{akai}
     H. Akai, and M. Ogura, Phys. Rev. Lett. {\bf 97}, 026401 (2006).
  \bibitem{xiao3}
     Y.-M. Nie, and X. Hu, Phys. Rev. Lett. {\bf 100}, 117203 (2008). 
  \bibitem{muller}
    G. M. M$\ddot{\mathrm{u}}$ller, J. Walowski, M. Djordjevic, G.-X. Miao, A. Gupta, A. V. Ramos, K. Gehrke, V. Moshnyaga, K. Samwer, J. Schmalhorst, A. Thomas, A. H$\ddot{\mathrm{u}}$tten, G. Reiss, J. S. Moodera, and M. M$\ddot{\mathrm{u}}$nzenberg, Nat. Mater. {\bf 8}, 56 (2009).  
  \bibitem{xiao4}
     S.-J. Hu, and X. Hu, J. Phys. Chem. C {\bf 114}, 11614 (2010). 
  \bibitem{coey}
     H. Kurt, K. Rode, P. Stamenov, M. Venkatesan, Y.-C. Lau, E. Fonda, and J. M. D. Coey, Phys. Rev. Lett. {\bf 112}, 027201 (2014).
  \bibitem{tokura}
    Y. Tokura, Rep. Prog. Phys. {\bf 69}, 797 (2006).  
  \bibitem{bgpark}
      B. G. Park, J. Wunderlich, X. Mart\'\i, V. Hol$\acute{y}$, Y. Kurosaki, M. Yamada, H. Yamamoto, A. Nishide, J. Hayakawa, H. Takahashi, A. B. Shick, and T. Jungwirth, Nat. Mater. {\bf 10}, 347 (2011).
      \bibitem{wang}
     Y. Y. Wang, C. Song, B. Cui, G. Y. Wang, F. Zeng, and F. Pan, Phys. Rev. Lett. {\bf 109}, 137201 (2012).
  \bibitem{bibes}
    D. Sando, A. Agbelele, D. Rahmedov, J. Liu, P. Rovillain, C. Toulouse, I. C. Infante, A. P. Pyatakov, S. Fusil, E. Jacquet, C. Carr$\acute{e}$t$\acute{e}$ro, C. Deranlot, S. Lisenkov, D. Wang, J-M. Le Breton, M. Cazayous, A. Sacuto, J. Juraszek, A. K. Zvezdin, , L. Bellaiche, B. Dkhil, A. Barth$\acute{e}$l$\acute{e}$my, and M. Bibes, Nat. Mater. {\bf 12}, 641 (2013).
  \bibitem{liang}
     Q.-F. Liang, L.-H. Wu, and X. Hu, New J. Phys. {\bf 15}, 063031 (2013). 
        \bibitem{mugavero}
  S. J. Mugavero III, A. H. Fox, M. D. Smith, and H.-C. zur Loye, J. Solid State Chem. {\bf 183}, 465 (2010). 
  \bibitem{kim1}
   B. J. Kim, H. Jin, S. J. Moon, J.-Y. Kim, B.-G. Park, C. S. Leem, J. Yu, T. W. Noh, C. Kim, S.-J. Oh, J.-H. Park, V. Durairaj, G. Cao, and E. Rotenberg, Phys. Rev. Lett. {\bf 101}, 076402 (2008). 
  \bibitem{kim2}
   B. J. Kim, H. Ohsumi, T. Komesu, S. Sakai, T. Morita, H. Takagi, and T. Arima, Science {\bf 323}, 1329 (2009). 
  \bibitem{okada}
    Y. Okada, D. Walkup, H. Lin, C. Dhital, T.-R. Chang, S. Khadka, W. Zhou, H.-T. Jeng, M. Paranjape, A. Bansil, Z. Wang, S. D. Wilson, and V. Madhavan, Nat. Mater. {\bf 12}, 707 (2013).   \bibitem{pesin}
  D. Pesin, and L. Balents, Nat. Phys. {\bf 6}, 376 (2010).   \bibitem{xiangang}
     X. Wan, A. M. Turner, A. Vishwanath, and S. Y. Savrasov, Phys. Rev. B {\bf 83}, 205101 (2011). 
  \bibitem{blaha}
P. Blaha, K. Schwarz, G. K. H. Madsen, D. Kvasnicka, and J. Luitz, {\it{WIEN2k, An  Augmented Plane  Wave+Local Orbitals  Program  for  Calculating  Crystal Properties}} (Technische Universit$\ddot{\mathrm{a}}$t Wien, Vienna, Austria, 2001), ISBN 3-9501031-1-2.
  \bibitem{perdew}
J. P. Perdew, K. Burke, and M. Ernzerhof, Phys. Rev. Lett. {\bf 77}, 3865 (1996).
  \bibitem{anisimov}
     A. I. Liechtenstein, V. I. Anisimov, and J. Zaanen, Phys. Rev. B {\bf 52}, R5467 (1995); V. I. Anisimov, F. Aryasetiawan, and A. I. Lichtenstein, J. Phys.: Condens. Matter {\bf 9}, 767 (1997).
  \bibitem{note}
	 The obtained results for the present materials are robust with $U=4-8$eV for Pr and $0.7-2$eV  for Ir, respectively \cite{anisimov}. The $U$ value of 1.25eV is chosen for Ir on the basis of a recent first-principles result which reproduces the experimental Mott-insulating state for an isovalent material La$_2$MgIrO$_6$ to the present material PMIO: G. Cao, A. Subedi, S. Calder, J.-Q. Yan, J. Yi, Z. Gai, L. Poudel, D. J. Singh, M. D. Lumsden, A. D. Christianson, B. C. Sales, and D. Mandrus, Phys. Rev. B {\bf 87}, 155136 (2013).
  \bibitem{kunes}
J. Kune\v{s}, P. Nov\'{a}k, R. Schmid, P. Blaha, and K. Schwarz, Phys. Rev. B {\bf 64}, 153102 (2001).
  \bibitem{kresse}
G. Kresse, and J. Furthm$\ddot{\mathrm{u}}$ller, Phys. Rev. B \textbf{54}, 11169 (1996).
  \bibitem{supp}
   	See Supplementary materials for details on calculation methods from VASP including a discussion on hopping integrals among $t_{2g}$ orbitals obtained by Wannier downfolding scheme, disorder effects and additional information on half metallicity for doping rate $x=0.5$, 1 and 1.5 in [Pr$_{2-x}$Sr$_x$MgIrO$_6$]$_2$.
  \bibitem{lee}
     K.-W. Lee, and W. E. Pickett, Phys. Rev. B {\bf 77},  115101 (2008).
  \bibitem{kittel}
	C. Kittel, {\it{Introduction to Solid State  Physics, 8$^{\rm{th}}$ ed.}} (Wiley, Hoboken, 2005).
  \bibitem{zaanen}
  J. Zaanen, G. A. Sawatzky, and J. W. Allen, Phys. Rev. Lett. {\bf 55}, 418 (1985).
  \bibitem{medling}
   S. Medling, Y. Lee, H. Zheng, J. F. Mitchell, J. W. Freeland, B. N. Harmon, and F. Bridges, Phys. Rev. Lett. {\bf 109}, 157204 (2012). 
  \bibitem{matt}
   L. F. Mattheiss, Phys. Rev. B {\bf 13}, 2433 (1976).
     \bibitem{onur}
     O. Erten, O. N. Meetei, A. Mukherjee, M. Randeria, N. Trivedi, and P. Woodward, Phys. Rev. B {\bf 87}, 165105 (2013).
\end{thebibliography}

\begin{thebibliography}{10}
  \bibitem{vasp} 
    G. Kresse, and J. Furthm$\ddot{\mathrm{u}}$ller, Phys. Rev. B \textbf{54}, 11169 (1996);  Comput. Mater. Sci. {\bf 6}, 15 (1996).
  \bibitem{wien}
    P. Blaha, K. Schwarz, G. K. H. Madsen, D. Kvasnicka, J. Luitz, $WIEN2k$, $An$ $ Augmented$ $Plane$  $Wave+Local$ $Orbitals$  $Program$  $for$  $Calculating$  $Crystal$ $Properties$ (Eds: K. Schwarz), Technische Universit$\ddot{\mathrm{a}}$t Wien, Vienna, Austria, 2001.
  \bibitem{pbe}
    J. P. Perdew, K. Burke, and M. Ernzerhof, Phys. Rev. Lett. \textbf{77}, 3865 (1996).
  \bibitem{Ueff}
    S. L. Dudarev, G. A. Botton, S. Y. Savrasov, C. J. Humphreys, and A. P. Sutton, Phys. Rev. B \textbf{57} 1505 (1998).
  \bibitem{MP}
    H. J. Monkhorst, and J. D. Pack, Phys. Rev. B \textbf{13}, 5188 (1976). 
\end{thebibliography}
\end{document}